\documentclass{article}
\usepackage{arxiv}

\usepackage[utf8]{inputenc} 
\usepackage[T1]{fontenc}    
\usepackage[hidelinks]{hyperref}       
\usepackage{url}            
\usepackage{booktabs}       
\usepackage[version=4]{mhchem}
\usepackage{amsfonts}       
\usepackage{mathtools}
\usepackage{nicefrac}       
\usepackage{microtype}      
\usepackage{lipsum}		
\usepackage{graphicx}
\usepackage[square,numbers]{natbib}
\usepackage{doi}
\usepackage{caption}
\usepackage{subcaption}
\usepackage{amsmath}

\usepackage{bm}
\renewcommand{\vec}[1]{\bm{#1}}
    
\usepackage{url}
\usepackage[dvipsnames]{xcolor}
\definecolor{newcolor}{rgb}{.8,.349,.1}

   \newcommand{\mean}[1]{\overline{#1}\,}
   \newcommand{\meank}[1]{\overline{#1}^k}
   \newcommand{\pmean}[1]{\overline{\overline{#1}}\,}
   \newcommand{\logmean}[1]{\overline{#1}^{\text{log}}}
   
   \newcommand{\hmean}[1]{\overline{#1}^{H}}

   \newcommand{\mC}{\mathcal{C}}
   
   \newcommand{\mP}{\mathcal{P}}
   \newcommand{\mF}{\mathcal{F}}
   
   \newcommand{\mI}{\mathcal{I}}

   \newcommand{\sfC}{{\sf C}}

\newcommand{\sfv}{{\sf v}}
\newcommand{\sff}{{\sf f}}

\newcommand{\dtp}{\delta^{\,+}}
\newcommand{\dtpk}{\delta^{\,+}_k}
\newcommand{\dtm}{\delta^{\,-}}
\newcommand{\dtmk}{\delta^{\,-}_k}
\newcommand{\dd}{\mathrm{d}}

   \usepackage{caption}
   \usepackage{subcaption}
   \usepackage{epsfig,epstopdf}
   \usepackage[normalem]{ulem}
   
\title{Entropy conservative discretization of compressible Euler equations with an arbitrary equation of state}%

\author{ \href{https://orcid.org/0009-0003-6376-768X}{\includegraphics[scale=0.06]{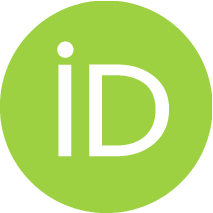}\hspace{1mm} Alessandro {Aiello}}\\
	Dipartimento di Ingegneria Industriale\\
	Universit\`a di Napoli ``Federico II''\\
	Napoli, Italy \\
	\texttt{alessandro.aiello@unina.it} \\
	\And
    \href{https://orcid.org/0000-0002-6518-3114}{\includegraphics[scale=0.06]{orcid.eps}\hspace{1mm} Carlo {De~Michele}}\\
	Unità Meccanica dei Fluidi\\
	Centro Italiano Ricerche Aerospaziali (CIRA)\\
	Capua, Italy \\
	\texttt{c.demichele@cira.it} \\
	\And
	\href{https://orcid.org/0000-0003-4943-9551}{\includegraphics[scale=0.06]{orcid.eps}\hspace{1mm}Gennaro Coppola} \\
	Dipartimento di Ingegneria Industriale\\
	Universit\`a di Napoli ``Federico II''\\
	Napoli, Italy \\
	\texttt{gcoppola@unina.it} \\
}

\renewcommand{\shorttitle}{EC discretization of compressible Euler equations with an arbitrary EoS}%

\hypersetup{
pdftitle={Entropy conservative discretization of compressible Euler equations with an arbitrary equation of state},
pdfsubject={physics.flu-dyn},
pdfauthor={Alessandro Aiello, Carlo De~Michele, Gennaro Coppola}
pdfkeywords={Compressible flows, Turbulent flows, Kinetic-energy preserving methods, Pressure equilibrium preserving method},
}

\begin{document}
\maketitle

\begin{abstract}
This study proposes a novel spatial discretization procedure for the compressible Euler equations which guarantees entropy conservation at a discrete level when an arbitrary equation of state is assumed. The proposed method, based on a locally-conservative discretization, guarantees also the spatial conservation of mass, momentum, and total energy and is kinetic-energy-preserving.
In order to achieve the entropy-conservation property for an arbitrary
non-ideal gas, a general strategy is adopted based on the manipulation of discrete balance equations through the imposition of global entropy conservation and the use of a summation by parts rule. The procedure, which is extended to an arbitrary order of accuracy, conducts to a general form of the internal-energy numerical flux which results in a nonlinear function of thermodynamic and dynamic variables and still admits the mass flux as a residual degree of freedom.  
The effectiveness of the novel entropy-conservative formulation is demonstrated through numerical tests making use of some of the most popular
cubic equations of state.
\end{abstract}

\keywords{Compressible flow \and Finite-volume \and Entropy conservation \and Real gases}

\section{Introduction} \label{sec:Introduction}
Numerical simulations are key in understanding the physical characteristics of turbulent compressible flows and are an invaluable tool for application-oriented studies~\cite{Pirozzoli_ARFM_2011}.
However, it is well known that, even in the absence of shock waves, standard discretizations are susceptible to strong nonlinear instabilities at high values of the Reynolds number, especially when non-dissipative schemes are employed for the discretization of the convective terms~\cite{Coppola_AMR_2019}. 
This behavior can be traced back to the failure of the numerical method to reproduce at a discrete level some structural properties or invariants of the continuous system of equations. 
In fact, the compressible Euler equations, which are a mathematical expression of the conservation of mass, momentum and total energy, have many induced symmetries which are typically not reproduced by the discrete system. 
The numerical community has progressively realized that the traditional focus on classical concepts of numerical analysis, such as order of accuracy and asymptotic convergence, has to be complemented by some more `physical' concepts, such as the discrete reproduction of the basic structural properties of the continuous formulation on an arbitrary finite mesh. 
These methods are referred to as \emph{structure preserving} or \emph{physics compatible} methods and have been the subject of many studies in recent years~\cite{Coppola_AMR_2019,Yee2020,Veldman_SIAMRev_2021,DeMichele_C&F_2023}.

Examples of structural properties for the compressible flow equations are the conservation of entropy (for smooth, nonviscous flows), the conservation of kinetic energy (for nonviscous flows in the incompressible limit), and the equilibrium of constant pressure and velocity distributions, which allows the existence of density waves moving at constant velocity and pressure.
Several numerical methods have been developed, which are able to preserve these properties at discrete level. 
Kinetic Energy Preserving (KEP) methods are the most well-known and studied among the others. These methods are designed to automatically preserve the conservative structure of the convective terms in the induced discrete equation for the evolution of kinetic energy. 
The study of KEP methods for compressible flows was pioneered by  \citet{Feiereisen_1981}.  Inspired by some formulations used for incompressible flows, the authors proposed for the first time the use of a skew-symmetric-like splitting of the convective terms in mass and momentum equations to induce a conservative structure of the convective terms in the kinetic-energy equation. This work stimulated the attention on this important topic and several other studies have appeared over the years, among which we mention here the Finite Volume (FV) analyses of \citet{Jameson_JSC_2008} and \citet{Subbareddy_JCP_2009} and the works on the triple splitting by \citet{Kennedy_JCP_2008} and \citet{Pirozzoli_JCP_2010}. 
More recently, also less conventional approaches have been explored, in which 
square-root variables are used \cite{Edoh_JCP_2022,Morinishi_JCP_2010,Rozema_JT_2014,DeMichele_AIAAAviation_2023}. A quite complete characterization of the possible KEP formulations in a Finite-Difference (FD) framework has been derived only recently~\cite{Coppola_JCP_2019}, and its relations with FV formulations have also been explored~\cite{Coppola_JCP_2023,Coppola_ECCOMAS_2022}. 
As KEP methods involve a proper discretization of mass and momentum equations alone, they are agnostic to the gas model used in the integration of the total-energy equation. KEP methods have been applied mainly in the context of ideal gases, but applications to more general equations of state have shown improved robustness as well \cite{Bernades_JCP_2023,Bernades_JSF_2024}. 

Discrete entropy conservation has also been explored to improve fidelity and suppress instabilities in turbulent simulations. 
Most of the contributions have been developed by following the elegant theory proposed by Tadmor~\cite{Tadmor_MC_1987,Tadmor_AN_2003} in the context of FV methods. 
Entropy Conservative (EC) numerical fluxes have been derived since the early papers on this subject in the context of ideal gases~\cite{Tadmor_MC_1987,Tadmor_AN_2003,Jameson_JSC_2008b}. 
\citet{Ismail_JCP_2009} developed `affordable' EC  fluxes based on the logarithmic mean, whereas \citet{Chandrashekar_CCP_2013} proposed for the first time both KEP and EC numerical fluxes.
More recently, Ranocha~\cite{Ranocha_JSC_2018,Ranocha_2020,Ranocha_CAMC_2021} proposed a KEP and EC formulation which is also able to enforce the so-called Pressure Equilibrium Preserving (PEP) property~(\cite{Shima_JCP_2021,Ranocha_CAMC_2021,Bernades_JCP_2023,DeMichele_JCP_2024}), i.e.~the ability of the discrete scheme to reproduce the traveling density wave solutions of the compressible Euler equations obtained for initially uniform pressure and velocity distributions.
Recently, \citet{DeMichele_JCP_2023} proposed a family of Asymptotically Entropy Conservative (AEC) schemes which are PEP and KEP, and arbitrarily reduce the error on entropy conservation. The main advantage of these schemes is that,  being based on arithmetic and harmonic means, they only use algebraic operations for the computation of the fluxes, and do not exhibit singularities, with increased efficiency at comparable performances.
AEC schemes improve already existing formulations which approximately enforce the entropy-conservation property, as the KEEP or KEEP$^{(N)}$ schemes~\cite{Kuya_JCP_2018,Tamaki_JCP_2022}.

An alternative approach to the theory of Tadmor to produce EC scheme is represented by the residual distribution schemes proposed by \citet{Abgrall2018}.
This correction procedure is not limited to the EC property, but it allows the incorporation of additional conservation laws into a baseline method. For example, it has been used to enforce various secondary properties such as angular momentum conservation \cite{Abgrall2022a} and KEP \cite{Abgrall2022}.
Furthermore, thermodynamically compatible schemes---having the entropy inequality as the primary evolution equation---have been produced which are able to discretely conserve total energy~\cite{Abgrall2023},
These approaches can be independent of the specific equation of state, however
a comparison with the proposed methodology will not be attempted in this study and is left for future research.

All the mentioned studies on formulations which are EC in the sense of Tadmor have been developed in the context of ideal gases, for which a simple linear relation between pressure and internal energy is assumed.
However, numerical simulations of real gases, often in transcritical regimes, have grown significantly in recent years, due to a wide range of possible engineering applications~\cite{Guardone_ARFM_2024,Muller_PoF_2016,Duronio_IJHMT_2024,Bernades_JSF_2024}.
Despite this growing interest, the literature on structure-preserving methods explicitly designed for real gases is still scarce. 
Most of the studies have been focused on the pressure equilibrium property, as the generation of spurious pressure oscillations, especially in the case of multi-component flows, is a longstanding issue in scale-resolving simulations with real gases~\cite{Abgrall_JCP_2001,Kawai_JCP_2015, Ma_JCP_2017,Bernades_JCP_2023}.
Despite these efforts, no numerical method is known at present that can discretely preserve the primary invariants mass, momentum, and total energy and simultaneously enforce the pressure equilibrium property or the conservation of entropy in the case of gases with an arbitrary equation of state.

In this paper, a general framework is developed to derive EC fluxes for the compressible Euler equations for real gases. The procedure is based on a conservative discretization of mass, momentum, and total-energy equations, which is also designed to discretely preserve the conservative structure of the convective terms in the induced evolution equation for the kinetic energy. As such, the resulting formulation spatially conserves the primary invariants, and the resulting fluxes are KEP and EC for gases with arbitrary equation of state. 
It is shown that a key element of the formulation is the internal-energy flux, which is a nonlinear function of the state variables that implicitly takes into account the nonlinear dependencies of the equation of state.
The treatment, which is firstly illustrated for second-order, two-point fluxes, is extended to an arbitrary order of accuracy and is shown to induce also entropy fluxes, resulting in an implicit locally-conservative discretization of the entropy equation.

A final note has to be made on the mathematical tools we use for our derivation. Entropy conservative discretizations of hyperbolic systems of conservation laws are typically investigated by using the theory of~\citet{Tadmor_MC_1987}, which relies on the specification of entropy variables and potentials and directly furnishes a constraint for the design of entropy fluxes. In this paper, we use a different approach essentially based on the enforcement of \emph{global} conservation of entropy, which is used to obtain constraints for the numerical fluxes of primary variables. Entropy fluxes are then derived using a quite general result on the equivalence between global and local conservation, giving an explicit decomposition of the induced entropy convective terms into a difference of numerical fluxes.
An analysis of the relation between these two approaches, which is worth investigating, is not attempted here and is left as a possible future research topic.

\section{Problem formulation} \label{sec:ProbForm}

The compressible Euler equations can be written as
\begin{align}
\dfrac{\partial \rho}{\partial t} &= -\dfrac{\partial \rho u_{\alpha}}{\partial x_{\alpha}} \;, \label{eq:Mass} \\[3pt]
\dfrac{\partial \rho u_{\beta}}{\partial t} &= -\dfrac{\partial \rho u_{\alpha}u_{\beta}}{\partial x_{\alpha}} -\dfrac{\partial p}{\partial x_{\beta}}  \;, \label{eq:Momentum} \\[3pt]
\dfrac{\partial \rho E}{\partial t} &=  -\dfrac{\partial \rho u_{\alpha}E}{\partial x_{\alpha}} -\dfrac{\partial p u_{\alpha}}{\partial x_{\alpha}},  \; \label{eq:TotEnergy}
\end{align}
where $\rho$ is the density, $p$ the pressure, $u_{\alpha}$ the Cartesian velocity component along the direction $x_{\alpha}$ and $E$ the total energy per unit mass, sum of internal and kinetic energies: $E = e + u_{\alpha}u_{\alpha}/2$. 
In this last equation, in Eqs.~\eqref{eq:Mass}--\eqref{eq:TotEnergy} and in what follows, the summation convention over repeated Greek indices is assumed.
The system \eqref{eq:Mass}--\eqref{eq:TotEnergy} is closed by the specification of an equation of state, which is assumed in the form $p = p(\rho,e)$ and is left as unspecified for the moment.

In this paper we focus on the discretization of the spatial terms of the system of equations~\eqref{eq:Mass}--\eqref{eq:TotEnergy} in the context of a semi-discretized procedure. As is usual in this framework, after the spatial discretization step has been performed, the resulting system of Ordinary Differential Equations (ODE) is integrated in time by using a standard solver, and we assume that the effects of time integration errors are negligible at sufficiently small time steps.
Consequently, in our theoretical analysis we assume that all the
manipulations involving time derivatives can be carried out at the continuous level.
Note that, as in many other similar studies, we deal with secondary conservation properties of spatial discretizations based on central schemes, which are specifically designed for shock-free simulations and, due to the absence of dissipation, are suited to turbulent flows in typically under-resolved simulations. 
Of course, these methods can also be used in smooth regions of shocked flows, in the spirit of hybrid methods~\cite{Pirozzoli_JCP_2002, Johnsen_JCP_2010}, which rely on shock sensors restricting the use of dissipative shock-capturing methods to regions near the shocks.

Although we will repeatedly refer to numerical fluxes, which are usually specified in the context of FV methods, our framework is mainly that of a FD treatment of the spatial terms in Eqs.~\eqref{eq:Mass}--\eqref{eq:TotEnergy}. 
Numerical fluxes arise in our analysis because we assume that the discrete formulation of the convective terms in Eqs.~\eqref{eq:Mass}--\eqref{eq:TotEnergy}, which are directly discretized, is always \emph{locally conservative}, i.e. it can be expressed as the sum of differences of numerical fluxes at adjacent faces along the three coordinate directions. A locally-conservative discretization is also \emph{globally conservative}, i.e.~the sum of all the pointwise contributions on the domain is zero (or a function of boundary conditions) by virtue of the telescoping property. These concepts are familiar in FV methods, but our numerical fluxes are always interpreted as functions of nodal values over a colocated uniform mesh, although a generalization of our findings to more general situations and to other discretization techniques can be easily carried out in most cases.
 Latin subscripts as $i,j$ or $k$ are used to denote the values of the discretized
variable on a nodal point $x_i$.

In addition to the balance equations for the primary variables mass, momentum and total energy, expressed by Eqs.~\eqref{eq:Mass}--\eqref{eq:TotEnergy}, one can also consider the induced equations for several kinematic and thermodynamic quantities of interest. 
By combining Eq.~\eqref{eq:Mass} and \eqref{eq:Momentum} one easily obtains the evolution equation for the specific kinetic energy $\rho\kappa = \rho u_{\alpha}u_{\alpha}/2$
\begin{equation}
\dfrac{\partial \rho \kappa}{\partial t} = -\dfrac{\partial \rho u_{\alpha} \kappa}{\partial x_{\alpha}} -u_{\alpha}\dfrac{\partial p}{\partial x_{\alpha}} \label{eq:KinEnergy} 
\end{equation}
and subtraction of Eq.~\eqref{eq:KinEnergy} from Eq.~\eqref{eq:TotEnergy} gives the evolution equation for internal energy
\begin{equation}
\dfrac{\partial \rho e}{\partial t} = -\dfrac{\partial \rho u_{\alpha} e}{\partial x_{\alpha}} -p\dfrac{\partial u_{\alpha}}{\partial x_{\alpha}}.\label{eq:IntEnergy}
\end{equation}
Finally, for adiabatic nonviscous flows the entropy of any particle of fluid remains constant during the motion, implying $D s/D t = 0$, where $s$ the specific entropy per unit mass and $D/Dt = \partial/\partial t + u_{\alpha}\partial/\partial x_{\alpha}$ is the material derivative. By using  Eq.~\eqref{eq:Mass} this condition leads to the balance equation for specific entropy:
\begin{equation}
\dfrac{\partial \rho s}{\partial t} = -\dfrac{\partial \rho u_{\alpha} s}{\partial x_{\alpha}}. \label{eq:entropy}
\end{equation}
Note that Eqs.~\eqref{eq:KinEnergy}--\eqref{eq:entropy} will be here considered as \emph{induced} equations, i.e. they do not constitute additional independent equations, once the primary Eqs.~\eqref{eq:Mass}--\eqref{eq:TotEnergy} have been assumed. Eqs.~\eqref{eq:KinEnergy}--\eqref{eq:entropy} can be easily derived from Eqs.~\eqref{eq:Mass}--\eqref{eq:TotEnergy} provided that the variables are smooth functions of $x_{\alpha}$ and $t$, in such a way that the basic rules of calculus are valid, and a suitable gas model has been specified. 
From a numerical point of view, a direct discretization of Eqs.~\eqref{eq:Mass}--\eqref{eq:TotEnergy}, even in a locally-conservative form,  does not guarantee, in general,  that the discrete evolutions of $\rho\kappa, \rho e$ and $\rho s$ follow discrete equations in which the convective terms are in locally- (or globally-) conservative form. 
General criteria for kinetic-energy preservation have been recently derived \cite{Jameson_JSC_2008,Coppola_JCP_2019,Veldman_JCP_2019,Coppola_JCP_2023} and can be used to guarantee also preservation of the internal energy~\cite{DeMichele_C&F_2023}. 
Entropy conservative methods, which are able to discretely mimic the conservative structure of Eq.~\eqref{eq:entropy}, have been developed for the special case of ideal gases~\cite{Chandrashekar_CCP_2013,Ranocha_2020}, for which there is a simple linear relation between pressure and internal energy, and entropy is defined as $s = c_v\log(p/\rho^{\gamma})$, where $c_v$ is the specific heat at constant volume and $\gamma$ is the ratio between the specific heats at constant pressure and volume. In the case of an arbitrary equation of state, no numerical method is known to discretely preserve the conservative structure of the induced balance equation \eqref{eq:entropy}. 
\section{Discrete formulation}
Eqs.~\eqref{eq:Mass}--\eqref{eq:entropy} have a common structure which can be expressed as
\begin{equation}\label{eq:GenStructBalEq}
    \dfrac{\partial \rho\phi}{\partial t}=-\mathcal{C}_{\rho\phi}-\mathcal{P}_{\rho\phi},
\end{equation}
where $\mathcal{C}_{\rho\phi}$ is the convective term
and $\mathcal{P}_{\rho\phi}$ is a pressure term.
In Eqs.~\eqref{eq:Mass}--\eqref{eq:entropy} $\phi$ assumes the values $1,u_{\beta},E,\kappa,e$ and $s$, respectively, whereas the pressure term is absent in mass \eqref{eq:Mass} and entropy \eqref{eq:entropy} equations.
The symbols $\mC$ and $\mP$ will be used indifferently to denote both the continuous or discrete versions of the spatial terms at the right-hand sides of Eqs.~\eqref{eq:Mass}--\eqref{eq:entropy}.
When needed, we will use the more explicit notation $\left.\mC_{\rho\phi}\right|_i$ to denote a particular discretization of $\mC_{\rho\phi}$ at node $x_i$.
The theory will be developed for the one-dimensional version of Eq.~\eqref{eq:Mass}--\eqref{eq:TotEnergy}, the extension to the three-dimensional case being straightforward, since the convective terms along the various directions are independent and can be treated separately by components. The 1D formulation uses an ordered set of 
grid nodes, 
thus we use a structured Cartesian mesh for 
the extension to two and three dimensions, and for carrying out the numerical tests (see Sec.~\ref{sec:NumResults_1D}). Such theoretical setting is, in principle, applicable also to unstructured meshes, at least for second-order, two-point fluxes. However, a high-order formulation as the one presented in Sec.~\ref{sec:HighOrder} should be redesigned due to the lack of ordering between the nodes if unstructured grids are used.
In this section and in the following Sec.~\ref{sec:ECFormulation} we will expose the theory with respect to second-order formulations, corresponding to two-point fluxes. In Sec.~\ref{sec:HighOrder} the extension to high-order fluxes for Cartesian meshes is carried on.

A locally-conservative discretization of the convective term $\mC_{\rho\phi}$ is one that can be recast as a difference of numerical fluxes at adjacent nodes along the relevant spatial direction. Using the symbol $\mF_{\rho\phi} = \left.\mF_{\rho\phi}\right|_{i+\frac{1}{2}} $ to denote the `right' numerical flux associated with the node $x_{i}$, one has 
\begin{equation}
\left.\mC_{\rho\phi}\right|_i = \dfrac{1}{h}\dtm \mF_{\rho\phi}\label{eq:DiffOfFluxes}
\end{equation}
where $h$ is the spatial step size and $\dtm $ is the \emph{backward} difference operator: $\dtm a_i = a_i - a_{i-1}$.
In what follows we will need to consider also the \emph{forward} difference operator, which is denoted by $\dtp$, with $\dtp b_i = b_{i+1}-b_i$.
The operators $\dtm$ and $-\dtp$ are adjoints with respect to the standard inner product.
This means that, neglecting the effects of boundary conditions (or by assuming periodic boundary conditions), the following property is easily seen to hold
\begin{equation}\label{eq:SBP_upwind}
    \sum_i b_i\dtm a_i = -\sum_ia_i\dtp b_i.
\end{equation}
Eq.~\eqref{eq:SBP_upwind} is a generalization of the Summation By Parts (SBP) property which is valid for central difference operators $\delta^0 = (\dtp +\dtm)/2$~\cite{Kravchenko_JCP_1997,Coppola_AMR_2019}. Expressed in matrix notation, the operators $\dtm$ and $\dtp $ correspond with the backward and forward bidiagonal difference matrices $D_{\text{bkw}}$ and $D_{\text{frw}}$ analyzed in the context of the `dual sided' discretizations in \cite{Coppola_JCP_2023} and  satisfying $D_{\text{bkw}}^T=-D_{\text{frw}}$, which is equivalent to Eq.~\eqref{eq:SBP_upwind}.

We will always assume that a KEP discretization is employed for mass and momentum equations~\cite{Coppola_AMR_2019,Coppola_JCP_2019}. 
In terms of numerical fluxes, KEP methods have the property that, for second-order discretizations, the convective fluxes for mass and momentum are linked by the relation $\mF_{\rho u} = \mF_{\rho}\mean{u}$ where $\mean{u} = (u_i+u_{i+1})/2$ is the arithmetic mean~\cite{Jameson_JSC_2008b,Veldman_JCP_2019}. 
It can be shown that, when a KEP discretization is employed, kinetic energy evolves according to a discrete equation analogous to Eq.~\eqref{eq:KinEnergy}, in which the convective term is in locally-conservative form,  with a numerical flux given, for second-order discretizations, by~\cite{Coppola_JCP_2023,DeMichele_C&F_2023}
\begin{equation}\label{eq:KinEnFlux}
    \mF_{\rho\kappa}=\mF_{\rho}\dfrac{u_iu_{i+1}}{2} .
\end{equation}
This observation induces us to formulate the numerical flux for total energy $\mF_{\rho E}$ as the sum of a suitably specified numerical flux for internal energy $\mF_{\rho e}$ and of the kinetic-energy flux induced by the KEP formulation, given by Eq.~\eqref{eq:KinEnFlux}.
In this case, the resulting formulation is equivalent, for exact temporal integration. to a formulation based on the direct discretization of the internal-energy equation~\eqref{eq:IntEnergy}, in place of Eq.~\eqref{eq:TotEnergy}. From now on, even if we numerically implement the discretization of the total-energy equation, we will focus on the specification of the internal-energy flux. The resulting total-energy convective flux is then built as the sum of the internal-energy flux and that of the kinetic energy induced by the KEP fluxes for mass and momentum in Eq.~\eqref{eq:KinEnFlux}. The conservative pressure term is correspondingly built as the sum of the discretized pressure terms in Eq.~\eqref{eq:KinEnergy} and \eqref{eq:IntEnergy}. 

\section{Entropy-conservative formulation}\label{sec:ECFormulation}
\subsection{General treatment}\label{sec:EC_GenTreat}
A KEP discretization of mass and momentum equations can be implemented with a latitude on the specification of the mass flux $\mF_{\rho}$. 
Once this has been done, the kinetic-energy flux is fixed as in Eq.~\eqref{eq:KinEnFlux}, and one is left with the internal-energy flux $\mF_{\rho e}$ as additional residual degree of freedom to satisfy further structural properties of the formulation. In this section, we analyze the induced discrete equation for the specific entropy $\rho s$ to obtain a formulation that is globally conservative of entropy. 

Our starting point is the Gibbs relation for a pure compound based on specific (per unit mass) variables 
in the form
\begin{equation}
    \dfrac{\partial e}{\partial t} = T\dfrac{\partial s}{\partial t} - p\dfrac{\partial v}{\partial t}
\end{equation}
where $v=1/\rho$ is the specific volume and $T = \left(\partial e/\partial s\right)_v$ is the temperature.
By manipulating time derivatives one easily arrives at the analogous equation valid for specific quantities per unit volume
 \begin{equation}\label{eq:Gibbs_t}
    \dfrac{\partial \rho e}{\partial t} = T\dfrac{\partial \rho s}{\partial t} + g\dfrac{\partial \rho}{\partial t}
\end{equation}
where $g = e-Ts+p/\rho$ is the Gibbs free energy.
This equation (also reported as Eq.~(15.5.2) in \cite{Kondepudi_2015}, p. 352)
reduces to $\partial_t (\rho s) = c_v/e \,\partial_t (\rho e) + (s-\gamma c_v)\,\partial_t \,\rho$ (see \cite[Eq.~(33)]{Coppola_JCP_2019}) in the case of ideal gases.
Equation~\eqref{eq:Gibbs_t} is now used to obtain a relation between the discretized spatial terms of mass, internal energy and entropy. 
In fact, substituting the right-hand sides of Eq.~\eqref{eq:Mass}, \eqref{eq:IntEnergy} and \eqref{eq:entropy}, as expressed in the generic form of Eq.~\eqref{eq:GenStructBalEq}, one has
\begin{equation}\label{eq:DiscrConvTerm_Rel}
    \mC_{\rho e} + \mP_{\rho e} = T\mC_{\rho s} + g\,\mC_{\rho} 
\end{equation}
We now assume that mass and internal-energy convective terms are discretized using a locally-conservative formulation, as expressed in Eq.~\eqref{eq:DiffOfFluxes}:
$\mC_{\rho} = \dtm\mF_{\rho}/h$ and $\mC_{\rho e} = \dtm\mF_{\rho e}/h$.
Moreover, 
for the pressure term in the internal-energy equation $\mP_{\rho e} = p\partial u/\partial x$ we use a standard, second-order central derivative formula, which can be expressed as 
\begin{equation}
    p\dfrac{\partial u}{\partial x}\simeq p\dfrac{\delta^0 u}{2h} = p\dfrac{\dtm \mean{u}}{h}.
\end{equation}
With these assumptions, Eq.~\eqref{eq:DiscrConvTerm_Rel} gives
\begin{equation}\label{eq:ConvEntropyTerm}
    \mC_{\rho s}\,h = \dfrac{1}{T}\,\dtm \mF_{\rho e} - \dfrac{g}{T}\,\dtm \mF_{\rho} + \dfrac{p}{T}\,\dtm \mean{u}
\end{equation}
Global conservation of entropy is now obtained by imposing that the sum of the convective terms over the domain is a function only of boundary terms, i.e. from Eq.~\eqref{eq:entropy}
\begin{equation}\label{eq:GlobalCons_continuous}
\dfrac{\dd}{\dd t}\int_{\Omega}\rho s\,\dd x = -\int_{\Omega}\mC_{\rho s}\,\dd x = \text{b.t.}  
\end{equation}
whose discrete version is
\begin{equation}\label{eq:GlobalCons}
    \sum_i\left.\mC_{\rho s}\right|_i\,h = \sum_i\left(\dfrac{1}{T}\,\dtm \mF_{\rho e} - \dfrac{g}{T}\,\dtm \mF_{\rho} + \dfrac{p}{T}\,\dtm \mean{u}\right) = \text{b.t.}
\end{equation}
Since we focus on the form of the fluxes for entropy conservation at interior points, we can 
neglect the effects of boundary conditions, which amounts to considering homogeneous or periodic boundary conditions. In this case, application of Eq.~\eqref{eq:SBP_upwind} furnishes
\begin{equation}
    \sum_i\left( \mF_{\rho e}\,\dtp\dfrac{1}{T}- \mF_{\rho}\,\dtp\dfrac{g}{T} + \mean{u}\,\dtp \dfrac{p}{T}\right) = 0
\end{equation}
for which a sufficient condition is that each term of the sum is individually zero, leading to 
\begin{equation}\label{eq:ConstrFluxes}
    \mF_{\rho e}\,\dtp\dfrac{1}{T}- \mF_{\rho}\,\dtp\dfrac{g}{T} + \mean{u}\,\dtp \dfrac{p}{T} = 0.
\end{equation}
Equation~\eqref{eq:ConstrFluxes} is the general constraint on the fluxes to nullify the spurious entropy production at interior points. Once one has assumed a suitable form for the mass flux $\mF_{\rho}$, it becomes a specification of the internal-energy flux
\begin{equation}\label{eq:ECFlux}
    \mF_{\rho e} = \mF_{\rho}\,\dfrac{\dtp g/T}{\dtp 1/T } - \mean{u}\,\dfrac{\dtp p/T}{\dtp 1/T}.
\end{equation}
Note that in the case of non-periodic boundary conditions, the present treatment still applies limited to interior points. In fact, application of the flux in  Eq.~\eqref{eq:ECFlux} at interior points guarantees that no spurious production of entropy is present in the interior of the domain. An \emph{ad hoc} treatment of boundary fluxes can be adopted to nullify possible spurious production of entropy at the boundaries.

Eq.~\eqref{eq:ECFlux} is the fundamental relation defining the class of EC schemes we analyze in this paper which still contains the mass flux $\mF_{\rho}$ as a residual degree of freedom, meaning that an arbitrary choice of the analytical flux discretization would still yield a valid EC scheme. In the present study we will only consider the scheme having as a mass flux simply the product of arithmetic averages of density and velocity: $\mF_{\rho}=\mean{\rho}\mean{u}$.
An investigation of other choices of $\mF_{\rho}$, all leading to KEP EC schemes that could be better optimized for further targets, is postponed to future research.

\subsection{Ideal gas}\label{sec:PerfectGas}
It is interesting to analyze how the constraint expressed in Eq.~\eqref{eq:ECFlux} simplifies in the special case of an ideal gas, for which $e =c_v T$ with a constant $c_v$, $p=\rho e (\gamma -1)$ and $s = c_v\log(p/\rho^\gamma)$. 
In this case, it is well known that  some EC formulations exist, which can also be KEP and Pressure-Equilibrium-Preserving (PEP).
Among the various options we consider here the  KEP, EC and PEP formulation proposed by \citet{Ranocha_JSC_2018} (see also \cite{Ranocha_arXiv_2017},\cite{Ranocha_2020} and \cite{Ranocha_CAMC_2021}), which is specified by the fluxes~\cite{DeMichele_JCP_2023}
\begin{equation}\label{eq:Ranocha_Flux_eint}
\mathcal{F}_{\rho}= \overline{\rho}^{\text{log}}\,\overline{u},\qquad\qquad
\mathcal{F}^*_{\rho u} =\mathcal{F}_{\rho}\,\mean{u} + \mean{p},\qquad\qquad
\mathcal{F}^*_{\rho E}=\mathcal{F}_{\rho}\,\left[\overline{\left(1/e\right)}^{\text{log}}\right]^{-1} + \mathcal{F}_{\rho}\,\frac{u_iu_{i+1}}{2} + \pmean{pu}
\end{equation}
where $\logmean{\phi} = \dtp\phi/\dtp\log\phi$ is the logarithmic mean, $\pmean{pu} = \left(p_iu_{i+1}+p_{i+1}u_i\right)/2$ is the product mean (corresponding to the FD approximation of $p\partial u/\partial x +u\partial p/\partial x$, representing the advective form of the pressure term in the total-energy equation), and the apex $^*$ is used to indicate that the flux includes also the contribution coming from the (conservative) pressure term.

In the ideal gas case, the Gibbs free energy is $g = e(\gamma-s/c_v) = c_vT(\gamma -\log p+\gamma\log\rho)$, and the internal-energy flux of Eq.~\eqref{eq:ECFlux} results in 
\begin{equation}\label{eq:ECFlux_ideal}
    \mF_{\rho e} = \gamma\mF_{\rho}\dfrac{\dtp\log\rho}{\dtp 1/e } - \mF_{\rho}\dfrac{\dtp \log p}{\dtp 1/e } - \mean{u}\dfrac{\dtp p/e}{\dtp 1/e}.
\end{equation}
In Eq.~\eqref{eq:ECFlux_ideal} the degree of freedom represented by the choice of the mass flux is still unconstrained. One possible use of that is to impose the PEP property. To do that, the internal-energy flux when velocity $U$ and pressure $P$ are constants should also be a constant dependent only on the values of $U$ and $P$ (see \cite{DeMichele_JCP_2024} for details). 
Indicating with a hat $\hat{\mF}$ the numerical flux evaluated at constant velocity and pressure, the PEP condition amounts to
\begin{equation}
    \hat{\mF}_{\rho e} = \dfrac{\gamma P}{\gamma-1}\hat{\mF}_{\rho}\dfrac{\dtp\log\rho}{\dtp \rho } - U P =\text{const.} \quad\implies \quad \hat{\mF}_{\rho}\dfrac{\dtp\log\rho}{\dtp \rho }=\text{const($U$,$P$)},
\end{equation}
which implies that imposition of the PEP property induces a constraint on the density flux. Assuming that it does not depend on pressure (\cite{Derigs_JCP_2017}),
consistency reduces this constraint to
\begin{equation}
    \hat{\mF}_{\rho} = \dfrac{\dtp \rho}{\dtp \log\rho} U = \logmean{\rho}U.
\end{equation}
There are multiple ways in which this constraint can be satisfied. A valid PEP and EC flux could be obtained even treating the product of $\rho$ and $u$ variables together, such as
\begin{equation}
    \mF_{\rho} = \dfrac{\dtp \rho u}{\dtp \log\rho u }.
\end{equation}
However, if we choose to treat $\rho$ and $u$ separately and to use for $u$ the same interpolation that we had in the pressure term of the internal-energy equation, then we have
\begin{equation}\label{eq:MassFluxRanocha}
    \mF_{\rho} = \logmean{\rho} \mean{u}
\end{equation}
which is the density flux in the discretization by Ranocha.
If we substitute this back in \eqref{eq:ECFlux_ideal} we obtain
\begin{multline}\label{eq:IntEnFluxRanocha}
    \mF_{\rho e} = \mF_{\rho}\left(\gamma\dfrac{\dtp\log\rho}{\dtp 1/e } - \dfrac{\dtp \log p}{\dtp 1/e } - \dfrac{\dtp\log\rho}{\dtp \rho } \dfrac{\dtp p/e}{\dtp 1/e}\right)=\\
   \mF_{\rho}\left(\gamma\dfrac{\dtp\log\rho}{\dtp 1/e } - \dfrac{\dtp \log p}{\dtp 1/e } - (\gamma-1)\dfrac{\dtp\log\rho}{\dtp \rho } \dfrac{\dtp \rho}{\dtp 1/e}\right)=\\
   \mF_{\rho}\left(- \dfrac{\dtp \log p}{\dtp 1/e } + \dfrac{\dtp\log\rho}{\dtp 1/e}\right)=
   \mF_{\rho}\left(\dfrac{\dtp\log(1/e)}{\dtp 1/e}\right)=\mathcal{F}_{\rho}\,\left[\overline{\left(1/e\right)}^{\text{log}}\right]^{-1}.
\end{multline}
which is the flux proposed by Ranocha for internal energy.
This shows that a KEP method with mass flux~\eqref{eq:MassFluxRanocha} and internal-energy flux given by Eq.~\eqref{eq:ECFlux} reduces to the Ranocha formulation \eqref{eq:Ranocha_Flux_eint} for ideal gases and is also EC and PEP.

\section{Entropy fluxes}

It is known that the scheme of Ranocha, specified by the fluxes in Eq.~\eqref{eq:Ranocha_Flux_eint},  is in locally-conservative form for the entropy equation, i.e. it admits induced entropy fluxes. 
In fact, at p. 26 of~\cite{Ranocha_arXiv_2017} (which is an extended version of \cite{Ranocha_JSC_2018}), the entropy numerical flux consistent with the formulation expressed in Eq.~\eqref{eq:Ranocha_Flux_eint}  is reported, and can be written, in our notation, as
\begin{equation}\label{eq:EntropyFluxRanocha}
    \mF_{\rho s} = \mF_{\rho}\left(\mean{s}-\gamma c_v+c_v\dfrac{\mean{e}}
    {\logmean{e}}\right)+c_v\left(\gamma -1\right)\mean{\rho}\mean{u},
\end{equation}
with $\mF_{\rho}=\logmean{\rho}\mean{u}$.

The scheme of Ranocha
has been derived in Sec.~\ref{sec:PerfectGas} as a special case, for ideal gases, of the novel and more general  formulation presented in Sec.~\ref{sec:EC_GenTreat} for gases with an arbitrary equation of state. 
However, this last formulation has been obtained by enforcing \emph{global} conservation of entropy, which does not imply, in general, that a \emph{local} entropy flux exists.
It is then legitimate to ask if the local conservation of entropy present in the ideal gas formulation is also present in the more general case.
To this aim, it is useful to recall here some results contained in 
 a recent paper by \citet{Coppola_JCP_2023}, where, in the context of an abstract theoretical analysis on the relations between the concepts of global and local conservation of primary and secondary invariants, a quite general equivalence between global and local conservation is established, at least in the framework of classical FD discretizations.
This result is used here to show that our formulation, obtained by imposing global conservation of entropy, admits local entropy fluxes for arbitrary gas models. Of course, these fluxes are consistent with the ones already known for the special case of ideal gases. 

A first observation contained in the paper by~\citet{Coppola_JCP_2023} which is relevant for us is that when the discrete convective term of the balance equation for a primary (i.e. directly discretized) or secondary (i.e. \emph{induced} in our terminology) variable can be expressed as the product of a matrix $\sfC$ times an arbitrary vector $\sfv$, global conservation is guaranteed when the matrix $\sfC$, which is in general a function of the state variables $\vec{w}$, has vanishing column sums.
To show this property, one  observes that a discrete version of the definite integral over the domain is obtained by premultiplication of the vector of discrete variables by the row vector $\mathbf{1}^{\text{T}} = \left[1,1,\ldots,1\right]$, from which one has the correspondence
\begin{equation}
    \begin{array}{c c c}
 \int\mC_{\rho\phi}\,\text{d}x =0& \quad\longrightarrow\quad & \mathbf{1}^{\text{T}}\sfC\sfv =0,
\end{array}
\end{equation}
which is a statement that discrete global conservation (cf. Eq.~\eqref{eq:GlobalCons_continuous}) corresponds with the vanishing of the column sums of the matrix $\sfC$.
The case of nonuniform mesh can also be included in this framework by taking into account the variable-size volumes of the cells into the time derivative term at the left-hand side of the balance equation \cite{Coppola_JCP_2023}.
The equivalence between global and local conservation follows then from a decomposition theorem valid for arbitrary vanishing column sums matrices, which allows one to split the vector $\sfC\sfv$ into the difference between a flux vector $\sff$ and its \emph{shifted} version $\widetilde{\sff}$, such that $\widetilde{\sff}_i=\sff_{i-1}.$
\begin{equation}
    \begin{array}{c c c}
 \mathbf{1}^{\text{T}}\sfC =\mathbf{0}^{\text{T}}& \quad\iff\quad & \sfC\sfv = \sff-\widetilde{\sff}.
\end{array}
\end{equation}
Moreover, the theorem reported in~\cite[Appendix~A1]{Coppola_JCP_2023} gives also an explicit formula for the derivation of the numerical flux $\sff$.
We will not go into the details of this derivation and refer the interested reader to the original paper and in particular to its Appendix A. 
Here we use this result to infer the existence of numerical entropy fluxes in all cases and obtain them with a more direct argument based on the derivation made in Sec.~\ref{sec:EC_GenTreat}.

We start by considering the expression for the entropy convective term in Eq.~\eqref{eq:ConvEntropyTerm} and rewrite it by making explicit the nodal dependencies and the difference operator $\dtm$
\begin{equation}\label{eq:ConvEntropyTerm_expl}
    h\left.\mC_{\rho s}\right|_i= \left(\dfrac{1}{T_i}\left.\mF_{\rho e}\right|_{i+\frac{1}{2}} - \dfrac{g_i}{T_i}\left.\mF_{\rho}\right|_{i+\frac{1}{2}} + \dfrac{p_i}{T_i} \mean{u}_i\right) - \left(\dfrac{1}{T_i}\left.\mF_{\rho e}\right|_{i-\frac{1}{2}} - \dfrac{g_i}{T_i}\left.\mF_{\rho}\right|_{i-\frac{1}{2}} + \dfrac{p_i}{T_i} \mean{u}_{i-1}\right).
\end{equation}
Now we observe that global conservation is guaranteed by Eq.~\eqref{eq:ConstrFluxes}, which also defines the numerical flux $\mF_{\rho e}$, and rewrite this equation again by making explicit the  nodal values and the difference operator $\dtp$
\begin{equation}\label{eq:ConstrFluxes_expl}
    \left(\left.\mF_{\rho e}\right|_{i+\frac{1}{2}}\dfrac{1}{T_{i+1}}- \left.\mF_{\rho}\right|_{i+\frac{1}{2}}\dfrac{g_{i+1}}{T_{i+1}} + \mean{u}_{i}\dfrac{p_{i+1}}{T_{i+1}}\right) -\left(\left.\mF_{\rho e}\right|_{i+\frac{1}{2}}\dfrac{1}{T_{i}}- \left.\mF_{\rho}\right|_{i+\frac{1}{2}}\dfrac{g_{i}}{T_{i}} + \mean{u}_{i}\dfrac{p_{i}}{T_{i}}\right) = 0
\end{equation}
from which one has the equality
\begin{equation}\label{eq:ConstrFluxes_expl2}
    \left(\left.\mF_{\rho e}\right|_{i+\frac{1}{2}}\dfrac{1}{T_{i+1}}- \left.\mF_{\rho}\right|_{i+\frac{1}{2}}\dfrac{g_{i+1}}{T_{i+1}} + \mean{u}_{i}\dfrac{p_{i+1}}{T_{i+1}}\right) = \left(\left.\mF_{\rho e}\right|_{i+\frac{1}{2}}\dfrac{1}{T_{i}}- \left.\mF_{\rho}\right|_{i+\frac{1}{2}}\dfrac{g_{i}}{T_{i}} + \mean{u}_{i}\dfrac{p_{i}}{T_{i}}\right).
\end{equation}
Observing that the right-hand side of Eq.~\eqref{eq:ConstrFluxes_expl2} is the same as the term in the first parentheses at the right-hand side of Eq.~\eqref{eq:ConvEntropyTerm_expl}, we can substitute the left-hand side of  Eq.~\eqref{eq:ConstrFluxes_expl2} into Eq.~\eqref{eq:ConvEntropyTerm_expl} to obtain
\begin{equation}\label{eq:ConvEntropyTerm_expl2}
    h\left.\mC_{\rho s}\right|_i= \left(\dfrac{1}{T_{i+1}}\left.\mF_{\rho e}\right|_{i+\frac{1}{2}}- \dfrac{g_{i+1}}{T_{i+1}}\left.\mF_{\rho}\right|_{i+\frac{1}{2}} + \dfrac{p_{i+1}}{T_{i+1}}\mean{u}_{i}\right) - \left(\dfrac{1}{T_i}\left.\mF_{\rho e}\right|_{i-\frac{1}{2}} - \dfrac{g_i}{T_i}\left.\mF_{\rho}\right|_{i-\frac{1}{2}} + \dfrac{p_i}{T_i} \mean{u}_{i-1}\right)
\end{equation}
which is in the form of a difference of fluxes, with numerical flux given by
\begin{equation}\label{eq:EntropyFlux_1}
    \left.\mF_{\rho s}\right|_{i+\frac{1}{2}}=\dfrac{1}{T_{i+1}}\left.\mF_{\rho e}\right|_{i+\frac{1}{2}}- \dfrac{g_{i+1}}{T_{i+1}}\left.\mF_{\rho}\right|_{i+\frac{1}{2}} + \dfrac{p_{i+1}}{T_{i+1}}\mean{u}_{i}.
\end{equation}
We finally note that, due the equality in  Eq.~\eqref{eq:ConstrFluxes_expl2}, the numerical flux in Eq.~\eqref{eq:EntropyFlux_1} could be expressed also as 
\begin{equation}\label{eq:EntropyFlux_2}
    \left.\mF_{\rho s}\right|_{i+\frac{1}{2}}=\dfrac{1}{T_{i}}\left.\mF_{\rho e}\right|_{i+\frac{1}{2}}- \dfrac{g_{i}}{T_{i}}\left.\mF_{\rho}\right|_{i+\frac{1}{2}} + \dfrac{p_{i}}{T_{i}}\mean{u}_{i}
\end{equation}
and any linear combination of Eqs.~\eqref{eq:EntropyFlux_1} and \eqref{eq:EntropyFlux_2} is also a consistent valid expression. 
Here we choose to use the more symmetric form given by the arithmetic average of Eqs.~\eqref{eq:EntropyFlux_1} and \eqref{eq:EntropyFlux_2}, leading to the final form
\begin{equation}\label{eq:EntropyFlux}
    \mF_{\rho s}=\mean{\left(\dfrac{1}{T}\right)}\mF_{\rho e}- \mean{\left(\dfrac{g}{T}\right)}\mF_{\rho} +\mean{\left(\dfrac{p}{T}\right)}\mean{u}.
\end{equation}

The flux in Eq.~\eqref{eq:EntropyFlux} is the entropy flux associated with the EC formulation presented in Sec.~\ref{sec:EC_GenTreat}. The internal-energy flux $\mF_{\rho e}$ is defined by Eq.~\eqref{eq:ECFlux} and the mass flux $\mF_{\rho}$ is still arbitrary and constitutes a degree of freedom which could be used to impart further structural properties to the formulation.
Substitution of Eq.~\eqref{eq:ECFlux} into Eq.~\eqref{eq:EntropyFlux} eventually furnishes
\begin{equation}\label{eq:EntropyFluxEntropyEq}
\mF_{\rho s} = \mF_{\rho}\left[\mean{\left(\dfrac{1}{T}\right)}
\dfrac{\dtp g/T}{\dtp 1/T}-\mean{\left(\dfrac{g}{T}\right)}\right]-
\mean{u}\left[\mean{\left(\dfrac{1}{T}\right)}
\dfrac{\dtp p/T}{\dtp 1/T}-\mean{\left(\dfrac{p}{T}\right)}
\right].
\end{equation}

To show that the flux in Eq.~\eqref{eq:EntropyFlux} and \eqref{eq:EntropyFluxEntropyEq} reduces to the entropy flux \eqref{eq:EntropyFluxRanocha} in the case of ideal gases, we can directly substitute the fluxes $\mF_{\rho}$ and $\mF_{\rho e}$ from Eq.~\eqref{eq:MassFluxRanocha} and \eqref{eq:IntEnFluxRanocha} into Eq.~\eqref{eq:EntropyFlux} and observe that in this case
\begin{equation}
    \mean{\left(\dfrac{1}{T}\right)} = c_v\mean{\left(\dfrac{1}{e}\right)},\qquad
    \mean{\left(\dfrac{g}{T}\right)} = c_v\gamma-\mean{s},\qquad
    \mean{\left(\dfrac{p}{T}\right)} = c_v\left(\gamma-1\right)\mean{\rho}
\end{equation}
from which Eq.~\eqref{eq:EntropyFluxRanocha} follows once one observes that $\mean{\left(1/e\right)}/\logmean{\left(1/e\right)}=\mean{e}/\logmean{e}$.

The determination of an entropy flux for the present formulation is useful if one wants to directly simulate the entropy equation, in place of the total-energy equation, and wants to preserve the total-energy balance. In fact, a direct discretization of Eq.~\eqref{eq:entropy} with numerical flux given by Eq.~\eqref{eq:EntropyFluxEntropyEq} gives a formulation which is equivalent, for exact time integration, to the formulation based on the discretization of the total energy with internal-energy flux given by Eq.~\eqref{eq:ECFlux}.

As a final comment, it is interesting to note the similarity between the formula of the numerical entropy flux presented in Eq.~\eqref{eq:EntropyFlux} and the one provided by the theory of \citet{Tadmor_MC_1987}, with 
the coefficients multiplying the numerical fluxes playing an analogous role to the entropy variables. 

\section{High-order formulation}\label{sec:HighOrder}

In deriving the  EC internal-energy flux in Eq.~\eqref{eq:ECFlux}, we implicitly assumed that the interpolations used for the mass flux $\mF_{\rho}$ and for the velocity in the pressure term $\mP_{\rho e}$ are two-point averages of the variables at the nodes $i$ and $i+1$ adjacent to the face $i+1/2$, thus implying a second-order accuracy for the whole procedure. 
However, the treatment can be extended to higher-order formulations by
combining two-point fluxes of variable width with suitable coefficients.
This approach has been used several times in the context of compressible Euler equations. \citet{LeFloch_SIAMJNA_2002} considered the two-point EC flux studied by Tadmor in a FV formulation, whereas \citet{Pirozzoli_JCP_2010} used a similar framework to extend to high order the bilinear fluxes used in FD contexts. \citet{Fisher_JCP_2013b} treated the case of non-periodic boundary conditions for the Tadmor flux, and \citet{Ranocha_JSC_2018} extended it to more general flux functions. We follow an adaptation of this last treatment contained in the Appendix A of the recent paper by \citet{DeMichele_JCP_2023}, which we recall here for convenience.

Given a two-point numerical flux $\mathcal{F}(\vec{w_i},\vec{w_{i+1}})$
for a generic quantity $\rho\phi$, which depends on the values of the variables vector $\vec{w}$ in the nodal points $i$ and $i+1$, we denote with  
$\mF^{i,k}=\mathcal{F}(\vec{w_i},\vec{w_{i+k}})$ its evaluation on the state variables at nodes $i$ and $i+k$.
Assuming that the numerical flux function is smooth with respect to its variables, symmetrical
(i.e. $\mathcal{F}(\vec{w_i},\vec{w_{i+k}}) = \mathcal{F}(\vec{w_{i+k},\vec{w_i}})$) and consistent with the continuous flux $f$, so that $\mathcal{F}(\vec{w_i},\vec{w_{i}}) = f(\vec{w_i})$, the high-order extension is built as 
\begin{equation}\label{eq:high_order_flux}
    \mathcal{F}^{h} = 2\sum_{k=1}^{L} a_{k}
    \sum_{m=0}^{k-1}\mF^{i-m,k},
\end{equation}
where the coefficients $a_k$ 
belong to a high-order central numerical derivative formula of the type 
\begin{equation}\label{eq:DerFormula}
    \left.\dfrac{\partial \phi}{\partial x}\right|_i \simeq\sum_{k = 1}^L\dfrac{a_k}{h}\delta_k \phi_i 
\end{equation}
with $\delta_k u_i = u_{i+k}-u_{i-k}$ and $L$ being the half-width of the stencil. If the numerical derivative formula in Eq.~\eqref{eq:DerFormula} has order $q$, the flux $\mF^h$ in Eq.~\eqref{eq:high_order_flux} has the same order of accuracy.
Using this theorem, a straightforward extension to high-order formulations can be accomplished by using Eq.~\eqref{eq:high_order_flux} for all the numerical convective fluxes involved in the procedure, namely that of mass, momentum and internal and kinetic energies. For the pressure terms, a similar treatment is used by observing that Eq.~\eqref{eq:DerFormula} 
can be rearranged as (cf. \cite{Pirozzoli_JCP_2010, Coppola_JCP_2019})
\begin{equation}\label{eq:DerFormula_I}
    \sum_{k = 1}^L\dfrac{a_k}{h}\delta_k \phi_i = \dfrac{1}{h}\dtm\left(2\sum_{k = 1}^La_k\sum_{m = 1}^{k-1}\meank{\phi}_{i-m}\right)
\end{equation}
where $\meank{\phi}_i = (\phi_{i+k}+\phi_{i-k})/2$ is a generalization of the arithmetic mean $\mean{\phi}$.
By denoting with $\mI_{\phi}$ the interpolation function in parentheses at the right-hand side of Eq.~\eqref{eq:DerFormula_I}, we have 
\begin{equation}
        \left.\dfrac{\partial \phi}{\partial x}\right|_i \simeq \dfrac{1}{h}\dtm\mI_{\phi}
\end{equation}
and the function $\mI_{\phi}$ plays the role of a high-order numerical flux for the terms expressed as a simple derivative. Its components $\mI_{\phi}^{i,k}$, which are analogous to $\mF^{i,k}$, are simply the arithmetic averages of the variables $\phi_i$ and $\phi_{i+k}$.

While this procedure guarantees a high-order discretization of the system of the Euler equations, a question arises on the EC character of the flux $\mF^h_{\rho e}$, since for the definition of the high-order version of the internal-energy flux, Eq.~\eqref{eq:high_order_flux} has to be used with $\mF_{\rho e}^{i,k}$ given by
\begin{equation}\label{eq:ECFlux_k}
    \mF_{\rho e}^{i,k} = \mF_{\rho}^{i,k}\dfrac{\dtpk g/T}{\dtpk 1/T } - \meank{u}\dfrac{\dtpk p/T}{\dtpk 1/T}
\end{equation}
and it is not clear a priori that the global flux satisfies the EC property.
To show that it is actually the case, we reconsider the derivation made in Sec.~\ref{sec:ECFormulation} and adapt it to the case of high-order fluxes.
Neglecting boundary terms,  Eq.~\eqref{eq:GlobalCons} becomes, 
\begin{equation}\label{eq:GlobalCons_H}
    \sum_i\left(\dfrac{1}{T}\dtm \mF^h_{\rho e} - \dfrac{g}{T}\dtm \mF^h_{\rho} + \dfrac{p}{T}\dtm \mI_{u}\right) = 0.
\end{equation}
Using Eqs.~\eqref{eq:high_order_flux} and \eqref{eq:DerFormula_I}, and by noting the equality
\begin{equation}
    \dtm\left(2\sum_{k = 1}^La_k\sum_{m = 1}^{k-1}\mF^{i-m,k}\right)=
    2\sum_{k = 1}^La_k\dtmk\mF^{i,k},
\end{equation}
Eq.~\eqref{eq:GlobalCons_H} reduces to
\begin{equation}
    \sum_i2\sum_{k=1}^La_k\left(\dfrac{1}{T}\dtmk \mF^{i,k}_{\rho e} - \dfrac{g}{T}\dtmk \mF^{i,k}_{\rho} + \dfrac{p}{T}\dtmk \meank{u}\right) = 0.
\end{equation}
Applying the SBP relation valid for $\dtpk$ and $\dtmk$
\begin{equation}\label{eq:SBP_upwind_k}
    \sum_i b_i\dtmk a_i = -\sum_ia_i\dtpk b_i
\end{equation}
one finally has 
\begin{equation}\label{eq:GlobalCons_Hk}
    2\sum_{k=1}^La_k\left[\sum_i\left(\mF^{i,k}_{\rho e}\dtpk\dfrac{1}{T} - \mF^{i,k}_{\rho}\dtpk\dfrac{g}{T} +  \meank{u}\dtpk\dfrac{p}{T}\right)\right] = 0
\end{equation}
for which the validity of Eq.~\eqref{eq:ECFlux_k} guarantees that each contribution in the inner parentheses vanishes.
This shows that if the high-order internal-energy flux is built according to Eq.~\eqref{eq:high_order_flux} with $\mF^{i,k}$ specified as in Eq.~\eqref{eq:ECFlux_k}, the global production of entropy is nullified, as in the second-order case.

\section{Numerical results}

In this section, the results of numerical simulations of the compressible Euler equations
are presented to verify the theoretical analysis and to assess the effectiveness of the newly proposed procedure. 
They are carried out for two cubic equations of state (Van der Waals and Peng-Robinson models) and for three configurations: a one-dimensional traveling density-wave test, a two-dimensional transcritical double shear layer, and the three-dimensional Taylor-Green vortex in supercritical conditions.
The balance equations for mass, momentum and total energy \eqref{eq:Mass}--\eqref{eq:TotEnergy} have been discretized by using a standard KEP formulation for mass and momentum, whereas the newly derived EC flux given in Eq.~\eqref{eq:ECFlux} has been used for the internal-energy contribution in the total-energy flux, resulting in the set of fluxes:
\begin{equation}\label{eq:Fluxes}
\mathcal{F}_{\rho}= \mean{\rho}\,\mean{u},\qquad\qquad
\mathcal{F}^*_{\rho u} =\mathcal{F}_{\rho}\,\mean{u} + \mean{p},\qquad\qquad
\mathcal{F}^*_{\rho E} = \mathcal{F}_{\rho e}^{EC} + \mathcal{F}_{\rho}\,\frac{u_iu_{i+1}}{2} + \pmean{pu}
\end{equation}
where $\mathcal{F}_{\rho e}^{EC}$ is the flux in Eq.~\eqref{eq:ECFlux}.
The formulation in Eq.~\eqref{eq:Fluxes} is second-order accurate; its extension to higher-order cases is made according to the theory in Sec.~\ref{sec:HighOrder}.
All the simulations have been carried out with a fourth-order 
explicit Runge-Kutta scheme for time integration,
except for the three-dimensional tests in Sec.~\ref{sec:NumResults_3D}, where a third-order, low-storage Runge-Kutta method is used.
The Courant number is taken sufficiently small in order to nullify the temporal errors on entropy conservation and focus solely on the spatial discretization errors. Of course, higher values of the Courant number can be used when such a specific need is not required.

As regards the singularity in the flux $\mathcal{F}_{\rho e}^{EC}$ defined in Eq.~\eqref{eq:ECFlux}, in all the tests we implement a simple fix consisting in the local use of a non-singular flux in regions in which the temperature distribution is nearly uniform.
In particular, the implementation is made by monitoring the local values of $\dtp T$. If this quantity is below a certain tolerance, we locally switch to a non-singular numerical flux. 
The threshold for the local values of $\dtp T$ is, at present, empirical and test-based, and will be declared for each numerical test shown. We found that a relatively high threshold value for $\dtp T$ does not spoil the EC property enforced by the novel formulation, whereas too small values could lead to losing precision in the evaluation of the fractions in \eqref{eq:ECFlux}, with the appearance of jumps in the temporal evolution of the global entropy-conservation error. 
The non-singular numerical fluxes used as a fix in this paper are the 
AEC schemes presented by \citet{DeMichele_JCP_2023}, which are designed to have excellent EC properties for ideal gases. 
This class of schemes, which are nonsingular and based on economic algebraic operations, constitutes a hierarchy of numerical fluxes based on a suitable truncated series expansion. They are designed to satisfy the KEP and PEP properties at each finite truncation, and asymptotically preserve the entropy, in the sense that the numerical error on global entropy production (in the ideal gas case) can be arbitrarily reduced by including additional terms in the expansion. 
In fact, taking two terms in the series expansion has been verified to be enough to avoid singularities without spoiling the EC properties of the new scheme.

Numerical tests have been carried out for molecular nitrogen $\ce{N2}$
near its critical conditions for both Van der Waals and Peng-Robinson Equations of State (EoS), whose structures are reported below. 
For each EoS, the thermodynamic state (e.g.~the state functions $e,s$) is evaluated by means of the suitable departure function, whose expressions are derived in \cite{Tosun}. Thus, in our framework, the differences of the thermodynamic exact differentials from an initial state $(T_1,p_1)$ to a final one $(T_2,p_2)$ are evaluated as 
\begin{align}
    \Delta e^{RG}_{1\rightarrow 2} &= -(e^{RG} - e^{IG})_{T_1,p_1} + \int_{T_1}^{T_2}c_v(T)\,dT + (e^{RG} - e^{IG})_{T_2,p_2}\\
    \Delta s^{RG}_{1 \rightarrow 2} &= -(s^{RG} - s^{IG})_{T_1,p_1} + \int_{T_1}^{T_2}\frac{c_p(T)}{T}\,dT - R_0\log{\frac{p_2}{p_1}} + (s^{RG} - s^{IG})_{T_2,p_2}
\end{align}
where the ideal-gas behavior, that is represented by the integral contributions, is given by the 5-coefficient polynomial fitting for $c_p(T)$ reported in \cite{NASA}. 

For comparison purposes, we selected \ce{N2}  for the simulations with both Van der Waals and Peng-Robinson models. As far as it concerns notation, we use $R_0$ to denote the universal gas constant whereas $\tilde{v}$ is the specific volume per unit mole, so that $\tilde{v} = \rho/M_w$, with $M_w$ the molecular weight. Finally, the subscript $(\cdot)_c$ indicates the critical conditions of the selected species.

The Van der Waals equation of state for pressure takes the form
    \begin{equation}
        p = \frac{R_0T}{\tilde{v} - b} - \frac{a}{\tilde{v}^2}
    \end{equation}
    with
    \begin{equation}    
        a = \frac{27}{64}\frac{R_0^2T_c^2}{p_c}, \qquad
        b = \frac{1}{8}\frac{R_0T_c}{p_c},
    \end{equation}
which can be seen as the simplest modification of the ideal-gas equation of state, taking into account the effective volume of molecules ($b$) and the interaction between the molecules themselves via the potential-energy-related term $a/\tilde{v}^2$. Despite its simplicity, such model remains widely used for highly-transcritical fluid flow simulations \cite{Pecnik}.
 On the other hand, the Peng-Robinson model \cite{PengRobinson} is
    \begin{equation} 
        p = \frac{R_0T}{\tilde{v}-b} - \frac{a(T)}{\tilde{v}^2 + 2b\tilde{v} - b^2}
    \end{equation}
    with
    \begin{equation}
        a(T) = 0.45724\frac{R_0^2T_c^2}{p_c}\bigg[1+c\bigg(1-\sqrt{\frac{T}{T_c}}\bigg)\bigg]^2\quad\text{ and }\quad
        b  = 0.07780\frac{R_0T_c}{p_c},
    \end{equation}
    where $c = 0.379642 + 1.48503\omega - 0.164423\omega^2 + 0.016666\omega^3$ for \ce{N2}, whose acentric factor $\omega$ is equal to $0.03720$.  Such model incorporates temperature dependency and the acentric factor inside the attractive term. The Peng-Robinson EoS has become one of the most popular  models to describe high-pressure/low-temperature fluids also in presence of interfaces \cite{Bernades_JCP_2023, Jofre_2021} due to its performances near the critical conditions.

The new formulation is compared with two schemes available in the literature, designed and used for ideal gases. 
The first one is the scheme proposed by Ranocha~\cite{Ranocha_JSC_2018,Ranocha_CAMC_2021}, already presented in Eq.~\eqref{eq:Ranocha_Flux_eint}. Since this scheme is KEP, EC and PEP for ideal gases, an evaluation of its performances on real gases can be highly informative of the effectiveness of the new formulation.
As a second reference scheme, we consider the formulation recently proposed by \citet{DeMichele_JCP_2023}, whose  fluxes are
\begin{equation}\label{eq:Flux_ArithHarmonic}
\mathcal{F}_{\rho}= \mean{\rho}\,\overline{u},\qquad\qquad
\mathcal{F}^*_{\rho u} =\mathcal{F}_{\rho}\,\mean{u} + \mean{p},\qquad\qquad
\mathcal{F}^*_{\rho E} = \mathcal{F}_{\rho}\hmean{e} + \mathcal{F}_{\rho}\,\frac{u_iu_{i+1}}{2} + \pmean{pu}.
\end{equation}
where $\hmean{\phi} = \phi_i\phi_{i+1}/(\phi_i+\phi_{i+1})$ is the harmonic mean. 
This set of fluxes constitutes the zeroth order case of the AEC family of schemes already illustrated, and will be here referred to as AEC${}^{(0)}$. Besides being KEP and PEP, it has shown remarkably good entropy conservation properties for the ideal gas case with respect to other classical schemes based on algebraic fluxes.

\begin{figure}
    \centering
\subfloat[Convergence analysis]{\includegraphics[width=0.45\linewidth]
{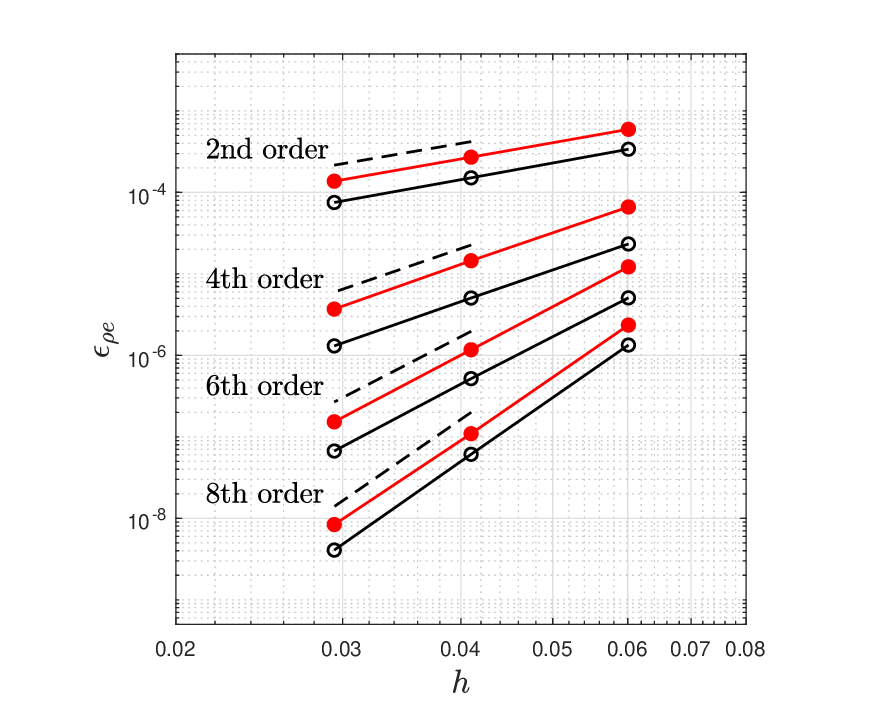}\label{fig:convergence_a}}\,
\subfloat[Entropy conservation]{\includegraphics[width = 0.45\linewidth]{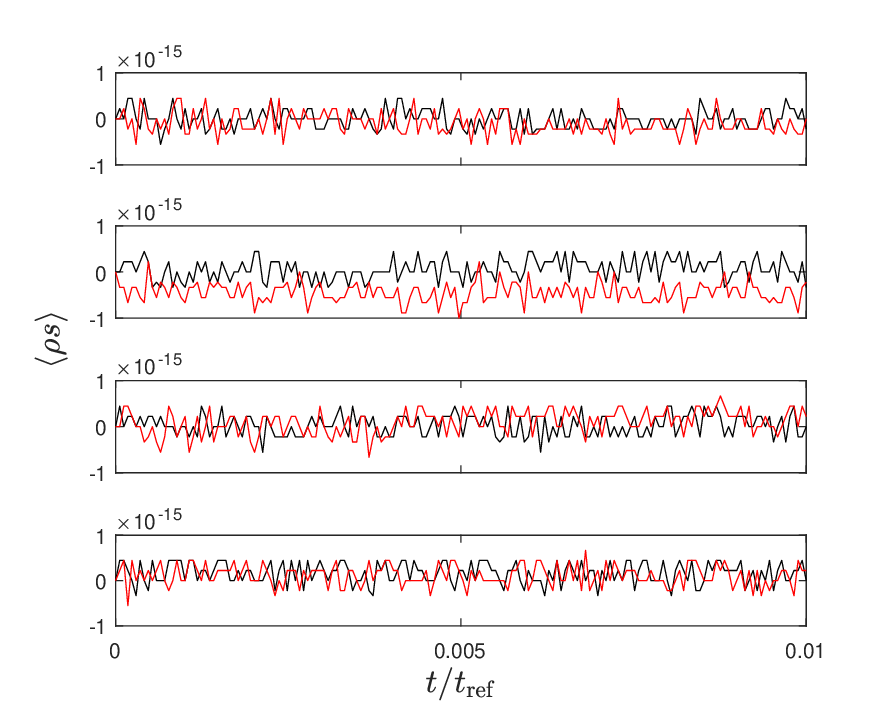}\label{fig:convergence_b}}
    \caption{Convergence analysis for high-order formulations. (a) Global error on internal energy as a function of the mesh size. Red lines with filled circles refer to the Van der Waals equation of state, black lines with empty circles refer to the Peng-Robinson model. (b) Entropy conservation error for $2$nd, $4$th, $6$th and $8$th order (from top to bottom) on a uniform mesh of $N=34$ intervals.}
    \label{fig:convergence}
\end{figure}
    
\subsection{1-D density wave}\label{sec:NumResults_1D}
The density-wave problem is a classical test case for compressible Euler equations, typically used to assess the 
PEP property.
In our framework, however, the density-wave test has been chosen as a simple one-dimensional case in which the entropy-conservation properties of the novel formulation can be assessed with minimum occurrences of locally uniform temperature distributions. This allows us to test the original fluxes by minimizing the use of the alternative non-singular formulations. In fact, for this particular test-case, the threshold value for switching to a non-singular formulation is set to $\dtp T < 10^{-14}$.

The simulation is carried out in the domain $[0,L]$, with $L=1\,\text{m}$
and periodic boundary conditions. The (dimensional) initial conditions are
\begin{equation}
    \begin{dcases}
        \rho(x,0) &= \rho_c\bigg[1 + \frac{1}{4}\sin{\bigg(\frac{n\pi}{L}x}\bigg)\bigg]\\
        u(x,0) & = u_0\\
        p(x,0) & = 2p_c
    \end{dcases}
\end{equation}
with $n=1$ and $u_0 = 10\,\text{m/s}$. 
We define a characteristic time $t_{\text{ref}} = L/u_{0}= 0.1\,\text{s}$,  indicating the time taken by the initial condition to travel across the whole domain. 

Preliminarily, in order to validate the high-order formulation presented in Sec.~\ref{sec:HighOrder}, and to test its entropy conservative character,  a convergence analysis is carried on.
The simulations are conducted until  $t/t_{\text{ref}}=0.01$ s with a Courant number based on $u_0$ equal to $0.002$.  The convergence is tested by calculating the normalized error $\epsilon_{\rho e}$ on internal energy given by the definition
\begin{equation}
    \epsilon_{\rho e} = \frac{\|\rho e - \widetilde{\rho e}\|_{\infty}}{\|\widetilde{\rho e}\|_{\infty}}
\end{equation}
where ${\rho e}$ is the numerical solution and $\widetilde{\rho e}$ is the analytical one evaluated on the grid points. 
To assess the EC character of the formulation, the global entropy production defined by
\begin{equation}
    \langle \rho s \rangle =
\dfrac{\int_{\Omega}  \rho s \,\dd \Omega - \int_{\Omega} \rho_0 s_0\, \dd \Omega}{ \left\lvert\int_{\Omega} \rho_0 s_0 \,\dd \Omega\right\rvert}
 \end{equation}
is monitored in time.
The results are reported in Fig.~\ref{fig:convergence}, where both the Van der Waals and the Peng-Robinson equations of state have been tested.
Fig.~\ref{fig:convergence_a} reports the convergence history for $N=16$, $24$ and $34$ intervals and high-order formulations up to the $8$th order. Fig~\ref{fig:convergence_b} reports the entropy production error for the case $N=34$ and for the various orders of accuracy, showing oscillations in the value of global entropy within machine precision with respect to its initial value, confirming the exact discrete conservation.
\begin{figure}
    \centering
    \subfloat[Van der Waals (2nd order)]{\includegraphics[width = 0.45\textwidth]{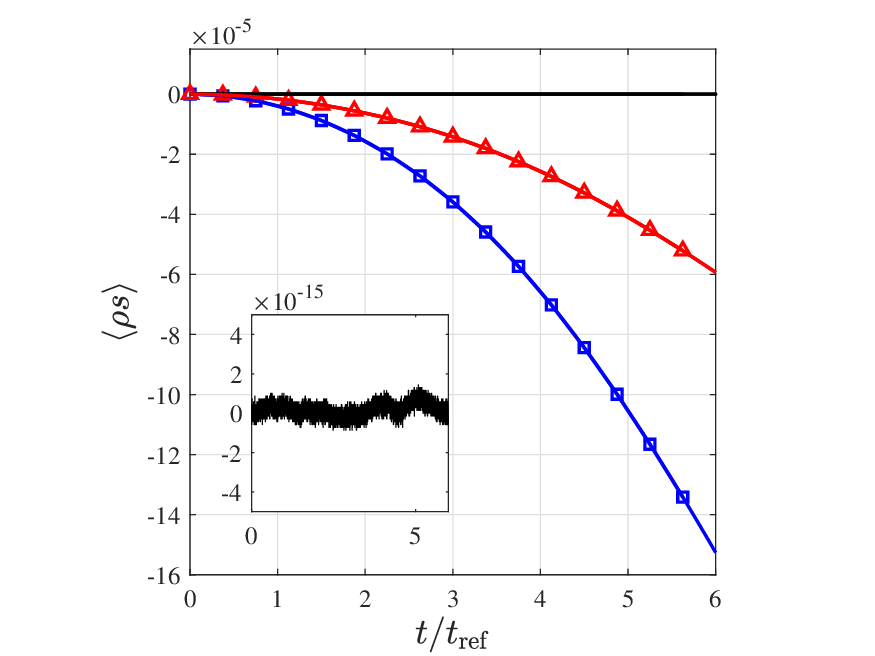}\label{fig:DW_a}}\,
    \subfloat[Peng-Robinson (4th order)]{\includegraphics[width = 0.45\textwidth]{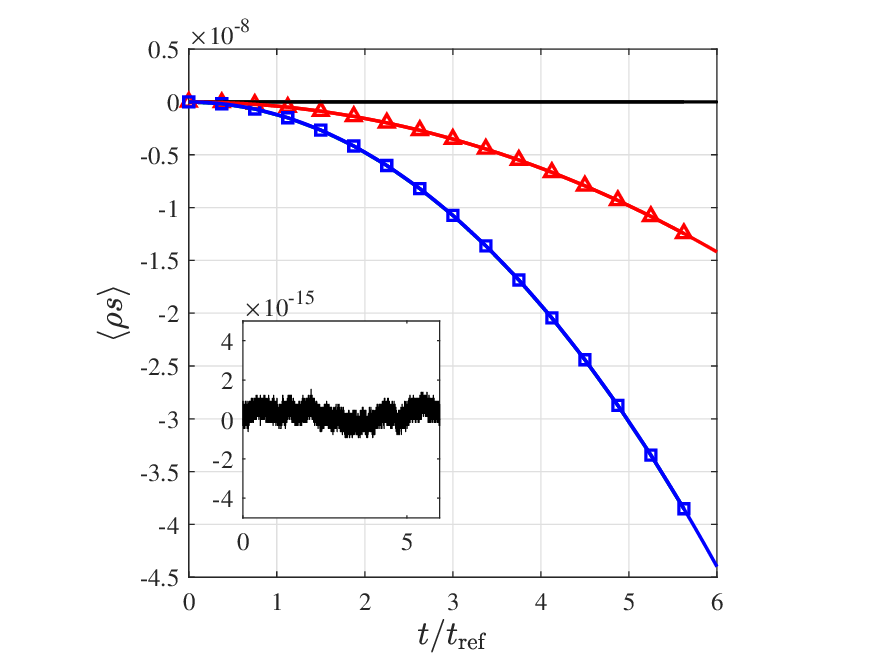}\label{fig:DW_b}}
    \caption{Entropy conservation in the 1-D density-wave test. Solid black lines represent the new EC scheme (which is also visualized in the inset), red lines with triangles represent the Ranocha scheme, blue lines with squares represent the AEC${}^{(0)}$ scheme.
    }
    \label{fig:DW}
\end{figure}

Fig.~\ref{fig:DW} shows the results of simulations with final time $t=0.6$ \text{s}
and $N=40$ intervals, with a Courant number set to $0.005$ for the Van der Waals (2nd order, Fig.~\ref{fig:DW_a}) case and $0.002$ for the Peng-Robinson (4th order, Fig.~\ref{fig:DW_b}) case, in order to be able to neglect time-integration errors.
The performances of the novel fluxes are compared with those of the Ranocha's and AEC${}^{(0)}$ formulations.

The insets clearly show that the new formulation reduces the global error on entropy production up to the machine precision, whereas the classical formulations exhibit a non-negligible spurious production of entropy which accumulates in time. As it is possible to note from Fig.~\ref{fig:DW}, the Ranocha scheme seems to perform better than the AEC${}^{(0)}$ scheme. The same simulations, not included here, have also been carried out with the AEC scheme with two additional terms of the series expansion which has been chosen as the non-singular flux to be used locally in the presence of singularities for the new EC flux. The results showed an essentially identical behavior to that of the scheme of Ranocha regarding 
the error $\langle \rho s\rangle$, which is in agreement with the theory presented in \cite{DeMichele_JCP_2023}.

\subsection{Transcritical double shear layer}
A two-dimensional double shear layer is simulated in the rectangle $[-L_x,L_x]\times[-L_y,L_y]$, with $L_x=0.5\,\text{m}$ and $L_y=0.25\,\text{m}$, discretized in $N_x\times N_y = 32\times 16$ intervals on a uniform mesh with periodic boundary conditions and   initial conditions
\begin{equation}
    \begin{dcases}
        u(x,y^\pm,0) &= u_0\bigg[1 \mp A\tanh{\bigg(\frac{y }{\delta}\bigg)}\bigg]\\
        v(x,y,0) & = \epsilon\sin{\bigg({\dfrac{k\pi}{L_x}}x\bigg)}e^{-4y^2/\delta}\\
        T(x,y^\pm,0) & = T_0\bigg[1 \pm B\tanh{\bigg(\frac{y}{\delta}\bigg)}\bigg]
    \end{dcases}
\end{equation}
where $A=B=3/8$, $\epsilon=0.1$, $\delta = 1/15$ and $k=3$. Courant number is set to $0.005$. The constants $u_0=20\,\text{m/s}$ and $\,T_0=110\,\text{K}$ are the streamwise component of the velocity and the temperature on the center line $(x,0,0)$, respectively. The $y-$ component of velocity, $v$, is initialized to provide a perturbation that triggers three roll-up vortices, while exponentially decaying with the distance from the centerline. Following \cite{Bernades_JCP_2023}, we set a bulk pressure $p = 2p_c$, whereas  the density $\rho$ is evaluated from the equation of state.
\begin{figure}
    \centering
    \subfloat[Van der Waals (4th order)]{\includegraphics[width =  0.45\textwidth]{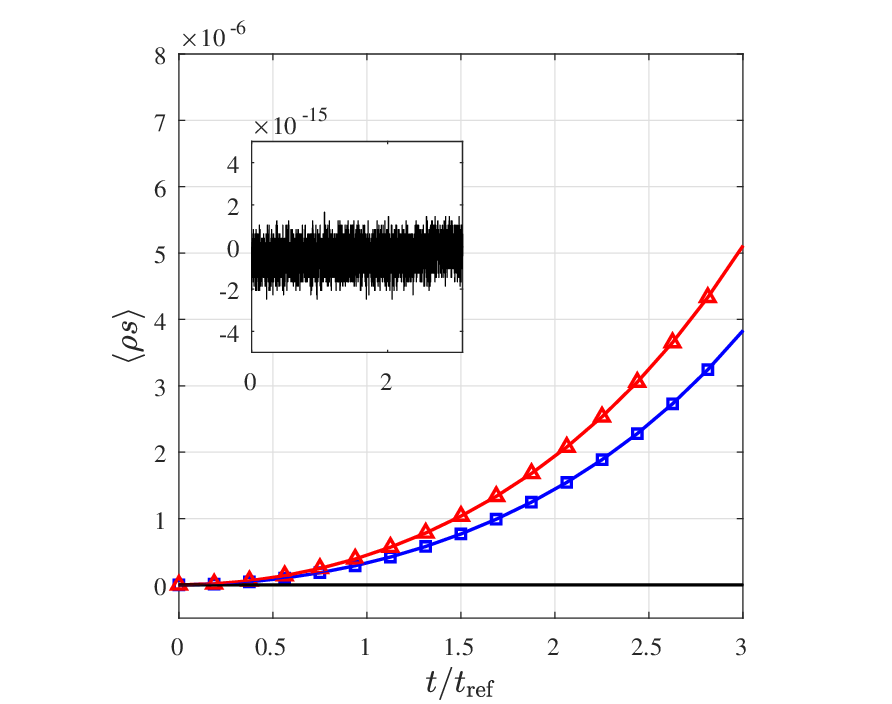}}\,
    \subfloat[Peng-Robinson (2nd order)]{\includegraphics[width = 0.45\textwidth]{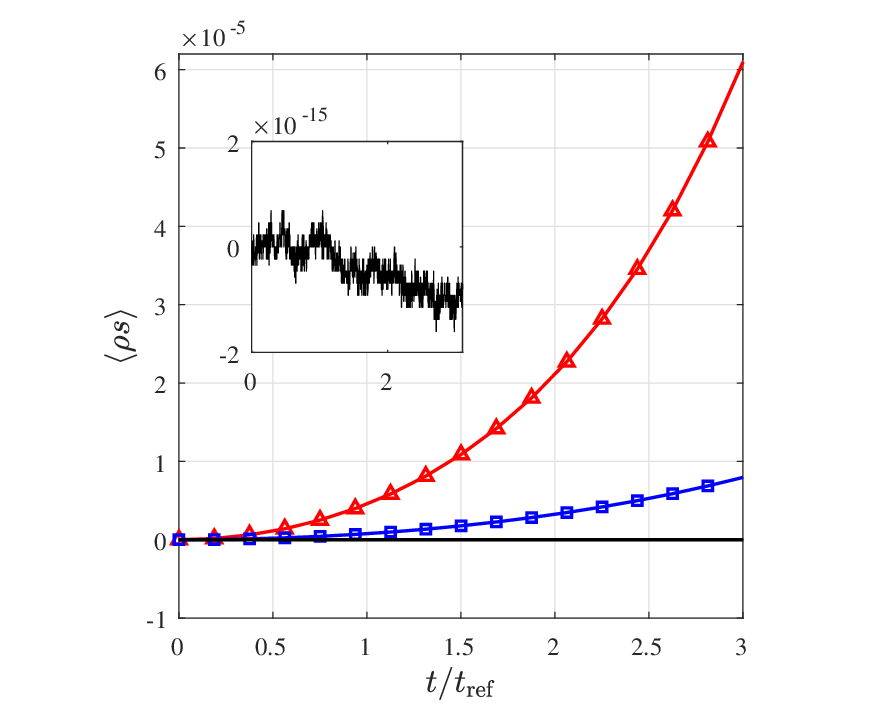}}
    \caption{Entropy conservation in the 2-D double shear layer. Solid black lines represent the new EC scheme (which is also visualized in the inset), red lines with triangles represent the Ranocha scheme, blue lines with squares represent the AEC${}^{(0)}$ scheme.
    }
    \label{fig:EC_mix_layer}
\end{figure}
The reference time $t_{\text{ref}}$ is defined by $ t_{\text{ref}}= \lambda/u_0$, where $\lambda = 1/k = 1/3\,\text{m}$, leading to the value  $t_{\text{ref}} = 0.017\,\text{s}$ (cf. \cite{Bernades_JCP_2023}).

From Fig.~\ref{fig:EC_mix_layer}, reporting the entropy production for the various schemes, it is seen that the theoretical conservation properties of the new formulation are confirmed also in this test. In this case, the harmonic numerical flux on internal energy provides a smaller error on the conservation of $\rho s$ if compared to the Ranocha scheme. In particular, the difference between the two discretizations appears to be more relevant for the Peng-Robinson equation of state. 
The threshold value for this test-case has been set to $\dtp T < 10^{-3}$.

\subsection{Supercritical Taylor-Green vortices}\label{sec:NumResults_3D}
The 3D Taylor-Green Vortex (TGV) test case is simulated in the tri-periodic box of size $L_x\times L_y\times L_z = (2\pi L)^3$, with $L=1\,\text{m}$, discretized in $N=32^3$ evenly-spaced points. The threshold value on temperature is set to be $\dtp T < 10^{-3}$. The simulations have been carried out in the open-source, scalable, GPU-accelerated code STREAmS-2.0 \cite{Bernardini2021, Bernardini2023}, which performs high-fidelity direct numerical simulations of high-speed compressible flows. 
The original code, which discretizes the compressible Navier-Stokes equations with the perfect gas EoS, has been adapted by including the Van Der Waals and Peng-Robinson models, as well as the newly derived fluxes for the convective terms given in Eq.~\eqref{eq:Fluxes}. The spatial integration is performed with a sixth-order central scheme, while the third-order, low-storage Wray's RK method already included in the original version of STREAmS-2.0, is used for the time-integration, with a $\text{CFL} = 0.001$ based on the speed of sound.

\begin{figure}
    \centering
    \subfloat[Entropy conservation]{\includegraphics[width=0.45\linewidth]{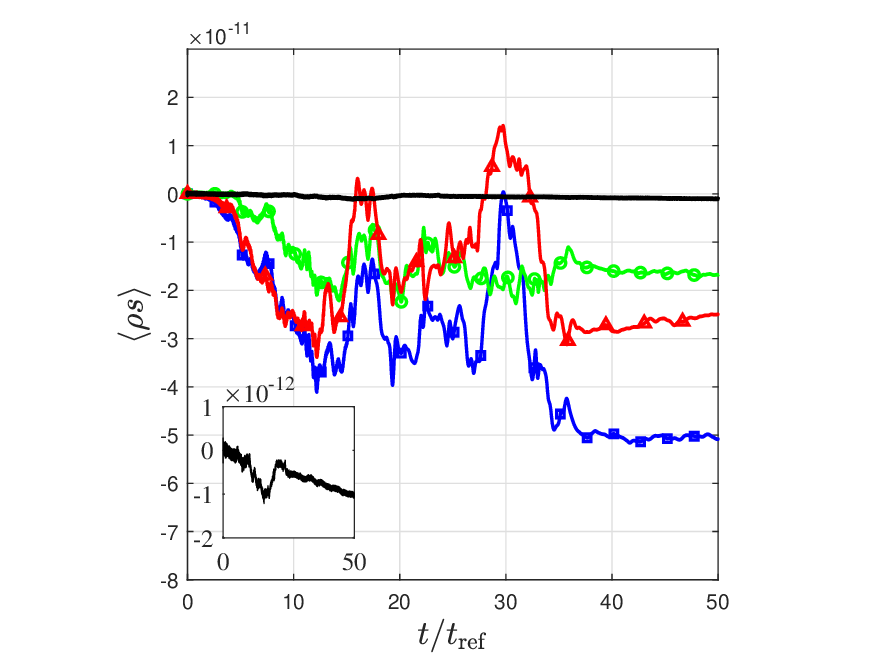}}\,
    \subfloat[Temperature fluctuations]{\includegraphics[width=0.45\linewidth]{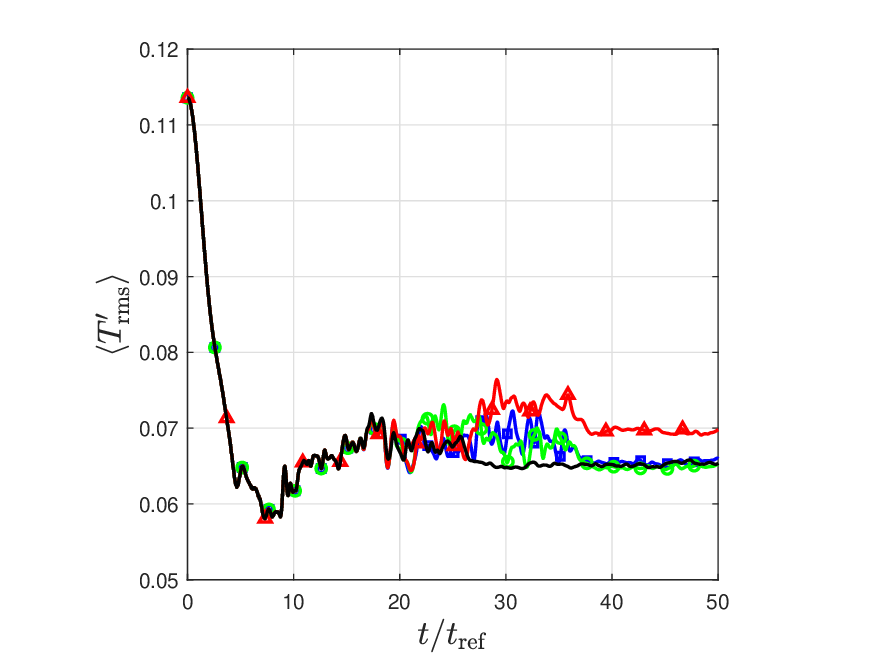}}\,
    \caption{Entropy conservation and thermodynamic fluctuations for the 3D TGV with the Peng-Robinson equation of state. Colors are set as in the previous figures, while green lines with circles represent the KEEP scheme, which is the built-in spatial discretization in STREAmS-2.0.}
    \label{fig:EC_TGV_PR}
\end{figure}

The initial conditions are set as
\begin{equation}
    \begin{cases}
        \rho(x,y,z,0) & = \rho_0\\
        u(x,y,z,0) & = u_0\sin{x}\cos{y}\cos{z}\\
        v(x,y,z,0) & = -u_0\cos{x}\sin{y}\cos{z}\\
        w(x,y,z,0) & = 0 \\
        p(x,y,z,0) & =  p_0 + \dfrac{\rho_0u_0^2}{16}(\cos{2x}+\cos{2y})(2+\cos{2z})
    \end{cases}
\end{equation}
with $u_0 = 20\,\text{m/s}$, $\rho_0 = 0.8\rho_c$, $p_0 = 10\,\text{MPa}$. The reference time $t_{\text{ref}}$ is defined as $L/u_0 = 0.05\,\text{s}$. The resulting temperature is $T\approx 185\,\text{K}$, so that the fluid is in supercritical conditions, and the subsequent Mach number is $M\approx 0.05$.
\begin{figure}
    \centering
    \subfloat[Entropy conservation]{\includegraphics[width=0.45\linewidth]{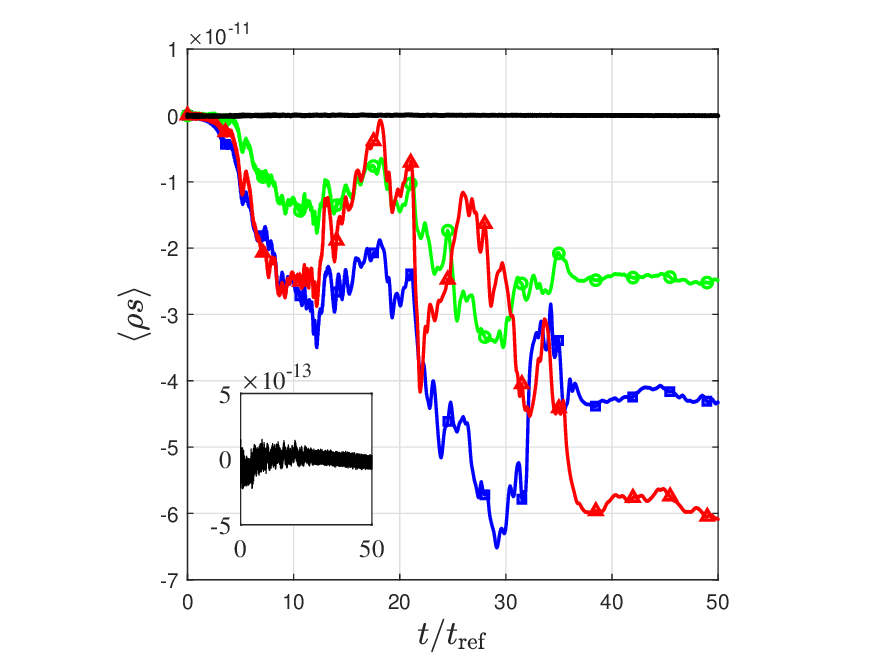}}\,
    \subfloat[Temperature fluctuations]{\includegraphics[width=0.45\linewidth]{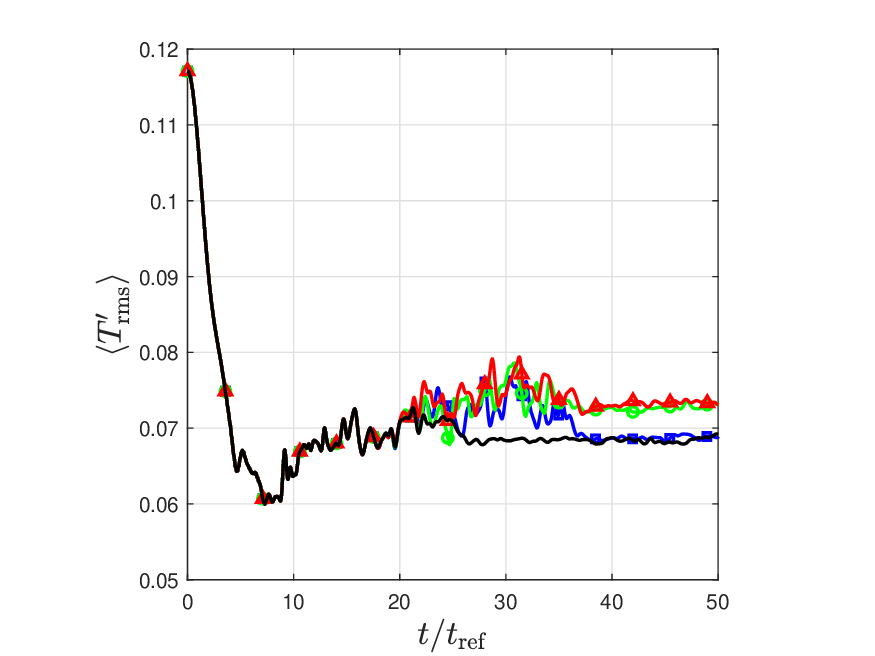}}\,
    \caption{Entropy conservation and thermodynamic fluctuations for the 3D TGV with the Van Der Waals equation of state. Colors are set as in the previous figures, while green lines with circles represent the KEEP scheme, which is the built-in spatial discretization in STREAmS-2.0.}
    \label{fig:EC_TGV_VDW}
\end{figure}
In Fig.~\ref{fig:EC_TGV_VDW}
and \ref{fig:EC_TGV_PR}  the entropy-conservation and the temperature fluctuations are shown for both the Van Der Waals and Peng-Robinson equations of state, respectively.
In addition to the Ranocha and AEC${}^{(0)}$ scheme already used in the previous test cases, we simulated the present test case by also using the KEEP schemes \cite{Kuya_JCP_2018}, already implemented in the STREAmS-2.0 code.
The proposed formulation consistently shows near machine-zero entropy conservation properties, although a slightly drifting behavior, which is attributed to the low-storage implementation of the third-order RK method, is visible for both the numerical tests. 
All the compared discretizations show that the fluctuations and the error on entropy conservation stabilize around an almost steady state after a transient behavior.
Density fluctuations (not shown here) have also been obtained, giving the same information provided by the temperature fluctuations.

\section{Conclusions}
In this study, we presented a general condition for the spatial discretization of the compressible Euler equations to obtain exact discrete conservation of entropy with an arbitrary equation of state, in addition to the conservation of primary invariants and kinetic energy preservation.
The proposed method, which was shown to be extendable to high-order formulations, has also been proven to be not only globally, but also locally conservative of entropy.
The conducted numerical tests verified the theoretical predictions on conservation properties and accuracy of the novel schemes for commonly employed cubic equations of state for real gas applications.

One potential drawback of the present formulation is that the produced numerical fluxes are potentially singular for certain uniform-flow conditions.
This peculiarity was already present in the EC formulations available for ideal gases. While this issue has already been addressed for these cases, a more general treatment that could avoid the local \emph{ad hoc} introduction of nonsingular fluxes adopted in this paper would be desirable and could constitute the subject of future refinements.

An additional element of novelty of the derivation presented in this paper is represented by the strategy used to obtain the new structure-preserving fluxes.
The additional conservation property is imposed in a fairly straightforward way, knowing the relation between the variable of interest and the set of primary (not necessarily conserved) variables whose equations are integrated.
This method has here been applied to entropy, but is believed to have a more general validity, constituting an alternative tool to the classical techniques used for similar problems.

The derived formulation is, to the best of our knowledge, the first entropy-conserving discretization of the compressible Euler equations specifically designed for real gases. Its structure-preserving properties make it an ideal candidate for high-fidelity simulations of turbulent flows where an arbitrary equation of state is used. Moreover, it constitutes the first building block toward a fully structure-preserving discretization of compressible flow equations for real gas applications.
In fact, the condition introduced in this work on the discretization of the convective term of internal energy introduces a whole family of EC schemes for which, however, no constraint is imposed on the choice of the numerical mass flux.
The use of this additional degree of freedom could be exploited to impart the scheme with supplementary structure-preserving properties, an example of which being the pressure equilibrium preservation property that is often desirable in the contexts of multi-component gases or multi-phase flows.
The exploration of this topic is left for future research.

\section*{Acknowledgments}
We acknowledge the CINECA award under the ISCRA initiative, for the availability of high-performance computing resources and support.

\vspace{1cm}

\bibliographystyle{model1-num-names}
\bibliography{Biblio_KEP_Compr}

\end{document}